\documentclass[letterpaper,10pt]{article} 
\usepackage{osameet3} %% use version 3 for proper copyright statement
\usepackage{siunitx} 
\usepackage{amsmath,amssymb}
\usepackage[colorlinks=true,bookmarks=false,citecolor=blue,urlcolor=blue]{hyperref} %pdflatex
\newcommand{\rfig}[1]{\normalsize Fig.~\ref{#1}}
\newcommand{\req}[1]{\normalsize Eq.~(\ref{#1})}

\usepackage{subcaption}
\long\def\/*#1*/{}

\begin{document}

\title{High-fidelity quantum transduction with long coherence time superconducting resonators}

\author{Changqing Wang, Silvia Zorzetti}
\address{Fermi National Accelerator Laboratory, PO BOX 500, Batavia, IL, USA}
\email{cqwang@fnal.gov, zorzetti@fnal.gov}
%%Uncomment the following line to override copyright year from the default current year.
%\copyrightyear{2022}

\begin{abstract}
\vspace{-0.001cm}

%\LaTeX{} users preparing manuscripts for Optica meetings or conferences
%should use  the \texttt{osameet3.sty} style file and should observe these
%guidelines to adhere to Optica requirements. Users of Bib\TeX{} may use the \texttt{osajnl.bst} style file, which is included in this distribution. Comments and questions should be directed to the Optica Conference Papers staff (tel: +1 202.416.6191,  e-mail: cstech@optica.org).

We propose a novel quantum transduction hybrid system based on the coupling of long-coherence time superconducting cavities with electro-optic resonators to achieve high-efficiency and high-fidelity in quantum communication protocols and quantum sensing.
\end{abstract}

\vspace{-0.0005cm}
\section{Introduction}
%Bulk Nb superconducting cavities are attractive for quantum applications for the high density of the electromagnetic field in the large microwave volume. These resonators are optimized for the operation in particle accelerators ($E_{\text max}\sim$\SI{10}{\mega\volta/\meter}). While they have the potential to enhance the coherence time when coupled to a qubit, as well as the conversion efficiency when hybridized with nonlinear materials. These devices would play a crucial role in the realization of long-distance quantum networks and in quantum sensing. 

High-efficiency and low-noise transduction in the quantum level remains challenging in the current designs and demonstrations. At Fermilab, we have developed bulk Nb superconducting radio-frequency (SRF) cavities with record-high 2 second photon lifetime, which represents a significant improvement compared to previous efforts in this field \cite{romanenko}. We are exploring hybrid coherent resonance systems and a bi-directional quantum transduction technology based on these high quality factor (Q) cavities to up/down- convert the information to/from the optical regime. The coupling between high-Q SRF cavities with nonlinear electro-optic resonators will herald a powerful quantum network. Moreover, such quantum systems with very low parasitic losses could exhibit maximized conversion efficiency at milli-Kelvin temperatures as well as high fidelity in quantum states transfer.
%The high density of electromagnetic field in the large microwave volume ($E_{\text max}\sim$\SI{10}{\mega\volta/\meter}) enhances the light-matter interaction and therefore the electro-optic effect.

%\section{Mcrowave-to-optical transduction approaches}
%Electro-optic modulators also appear to be promising components for the coupling of microwave and optical fields. Nowadays, the interest for these mediators is increasing as the carrier is directly modulated by the electric field by self-mixing in nonlinear devices, therefore with lower added noise and thermal quasiparticle poising. 

%This research focuses on \textbf{hybrid systems based on high-Q SRF cavities and electro-optic converters}. In the next sections this design choice will be motivated.

%\vspace{-0.5cm}
\section{Hybrid Quantum Systems for quantum transduction}
Non-centrosymmetric crystals are used to create interactions between microwave and optical fields using photonic RF three-wave-mixing processes. An electric field applied on the material modulates the refractive index and the incident optical field linearly with the RF voltage ($\chi^{(2)}$), known as the Pockels effect \cite{levi}. 
%In transduction, the self-heterodyne techniques have the advantage of eliminating the local oscillator (LO), used in conventional super-heterodyne mixers, reducing therefore complexity, power consumption, added noise, and thermal quasiparticles poisoning. 
We design a 3D hybrid system for quantum transduction through the integration of bulk Nb SRF cavities with optical resonators made of Lithium Niobate (LN: $\text L\text i\text N\text b\text O_\text3$), which is among the best electro-optic materials, with a large electro-optic coefficient ($r_{33}=\SI{31}{\pico\meter/\volt}$ at \SI{9}{\giga\hertz}) and low optical loss. These crystals exhibit low noise and high flexibility in mechanical designs to mitigate piezoelectric coupling with the microwave fields; dielectric losses are also well mitigated in large RF volumes, as demonstrated in Quantum Information Science (QIS) applications.

To describe the cavity electro-optic interaction, we use a triple-resonance scheme, in which the pump optical mode ($p$) is driven to coherently couple the optical signal mode ($a$) with the microwave mode ($b$), with the electro-optic (EO) coupling strength $g_{eo}$ \cite{HQS_girvin}. The Hamiltonian to model this system can be reduced to the following form:
\begin{equation}
    \hat{H}= \hslash g_{eo} (pa^\dag + p^\dag a) (b+b^\dag).
\end{equation}

The bidirectional conversion efficiency ($\eta$) depends on both the conversion and the coupling losses:
\begin{equation}\label{eq_efficiency}
    \eta=\frac{\kappa_{a,ex}}{\kappa_a}\frac{\kappa_{b,ex}}{\kappa_b}\times\frac{4C}{(1+C)^2}, \hspace{1cm}C=\frac{4n_p{g}_{eo}^2}{\kappa_a\kappa_b}
\end{equation}
where, for a generic $m$ mode ($m=a,b$), $\kappa_{m}$ is the total loss rate, consisting of both the intrinsic and external losses: $\kappa_m=\kappa_{m,ex}+\kappa_{m,i}$. $C$ is the cooperativity between the optical and the microwave modes, which depends on the photon number in the pump mode ($n_p$). The second term of \req{eq_efficiency} is the internal efficiency: $\eta_i=\frac{4C}{(1+C)^2}$.\\
%\begin{equation}\label{cavity_cooperativy}
%    C=\frac{4n_p{g^2}_{eo}}{\kappa_a\kappa_b}.
%\end{equation}

%\subsection{Operation mode}

The mentioned transduction approach requires the presence of a strong laser field that is used to bridge the energy gap between optical and microwave excitations. 
The operation mode is defined by the laser pump detuning. In a \textit{Red-detuned} scheme, the laser is tuned below the optical resonant frequency. This is an ideal operation mode for quantum transducers to achieve high conversion efficiency with low noise. Microwave and optical photons are converted via a beam splitter interaction. Maximum (unitary) conversion efficiency ($C=1$) is achieved at the critical coupling: $4n_p{g}_{eo}^2=\kappa_a \kappa_b$, in the ideal case of zero parasitic losses. When \textit{Blue-detuned}, the laser is tuned above the optical resonant frequency. A pair of microwave and optical photons are created with non-classical correlation between them. This two-mode squeezing process can be utilized to create entanglement between remote quantum systems, whereas the transduction efficiency is low ($\eta<<1$) \cite{tsang, Loncar}. However low losses in long coherence time quantum memories can be leveraged to enhance the transduction efficiency while keeping high the entanglement fidelity in quantum networks. The effective coupling strength ($g_{eo}^2n_p$) and the conversion efficiency is proportional to the laser pump power, which typically ranges from fractions to few mW. If the dissipated power is too high, thermal photons will be produced and degrade the superconductivity of the cavity along with the fidelity of the transduction. Another figure of merit that is of great importance in most quantum transducers' applications is the conversion efficiency, as defined in \req{eq_efficiency}, i.e. the success probability of the transduction process. While several parameters are impacting on the conversion efficiency, here we focus on the microwave photon lifetime, i.e., the coherence time of the microwave cavity.

%\begin{description}
%  \setlength\itemsep{0em}

%\item [Red-detuned:] the laser is tuned below the optical resonant frequency. This is an ideal operation mode for quantum sensors to achieve high conversion efficiency. Microwave and optical photons are converted via a classical beam-splitter interaction. It is possible to achieve maximum (unitary) conversion efficiency when $C=1$ and the critical coupling ($4n_p{g^2}_{eo}=\kappa_a \kappa_b$), in the ideal case of zero parasitic losses \cite{tsang}. 
%\item[Blue-detuned:] the laser is tuned above the optical resonant frequency. A non-classical correlation between the microwave and optical photons is created. A pair of photons is generated and can be entangled with remote quantum systems. This could be a low-efficiency process ($C<<1$), however low losses in long coherence quantum memories can play a role to enhance the fidelity in quantum communication \cite{tsang, Loncar}.
%\end{description}

\section{High fidelity quantum state transfer}

\begin{figure}[ht!]
   \centering
  \begin{subfigure}{0.32\textwidth}
 \includegraphics[width=\textwidth]{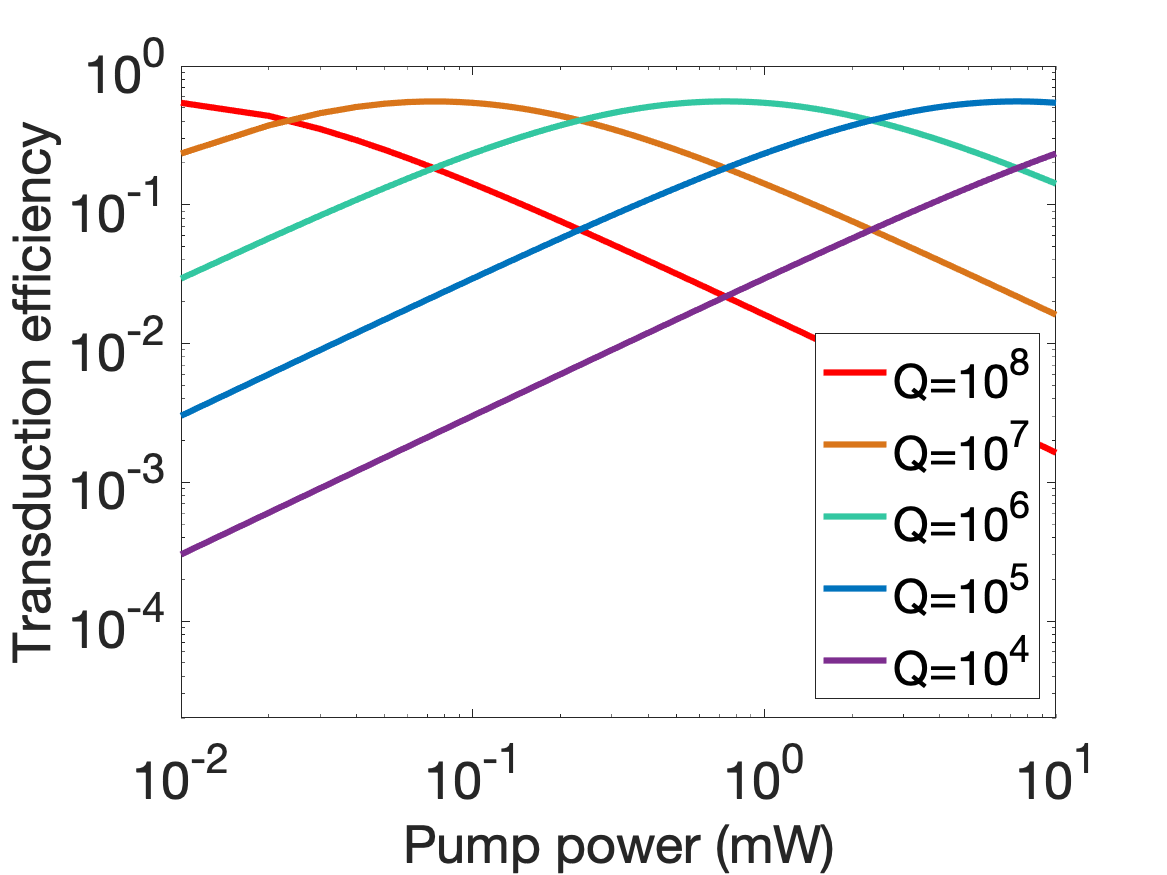}
    \caption{}
   \label{fig:transduction_efficiency}
    \vspace{-0.5cm}
  \end{subfigure}
      \begin{subfigure}{0.32\textwidth}
   \includegraphics[width=\textwidth]{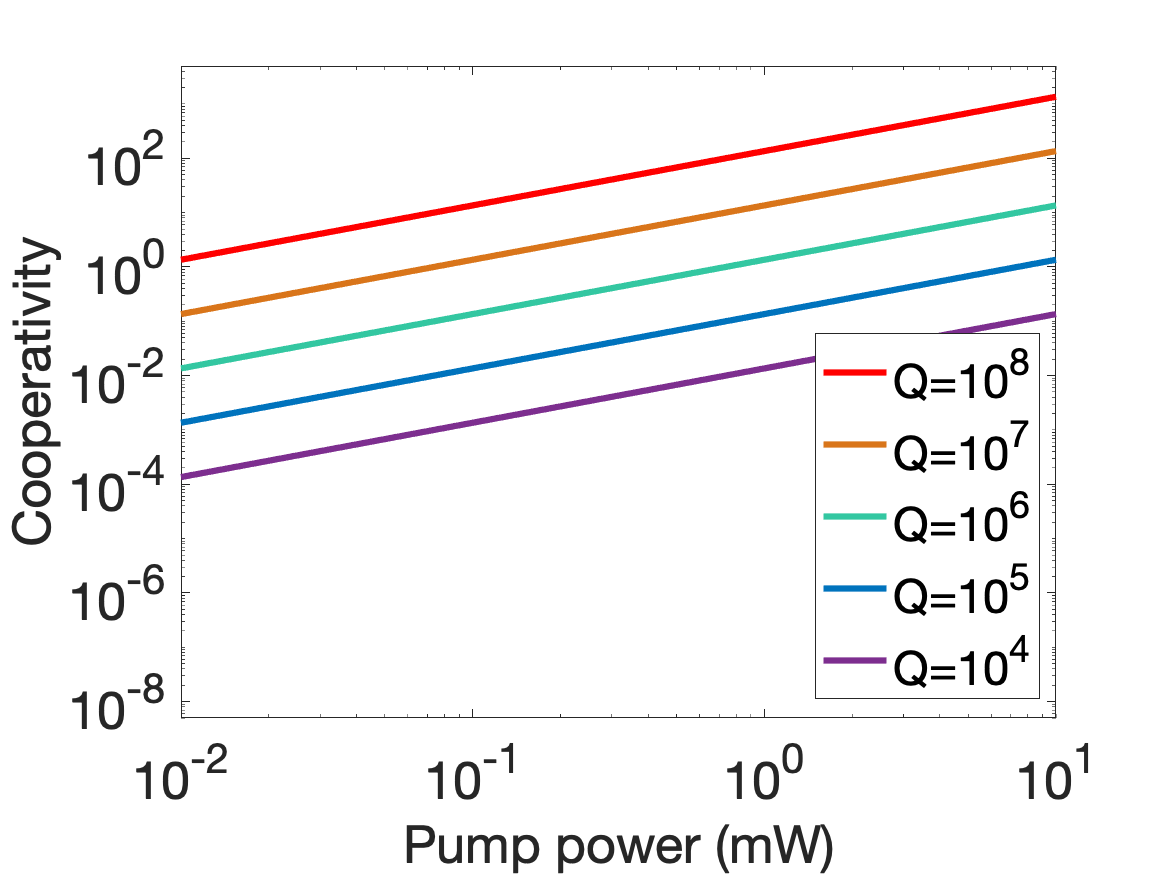}
  \caption{}
 \label{fig:cooperativity}
  \vspace{-0.5cm}
 \end{subfigure}
    \begin{subfigure}{0.32\textwidth}
   \includegraphics[width=\textwidth]{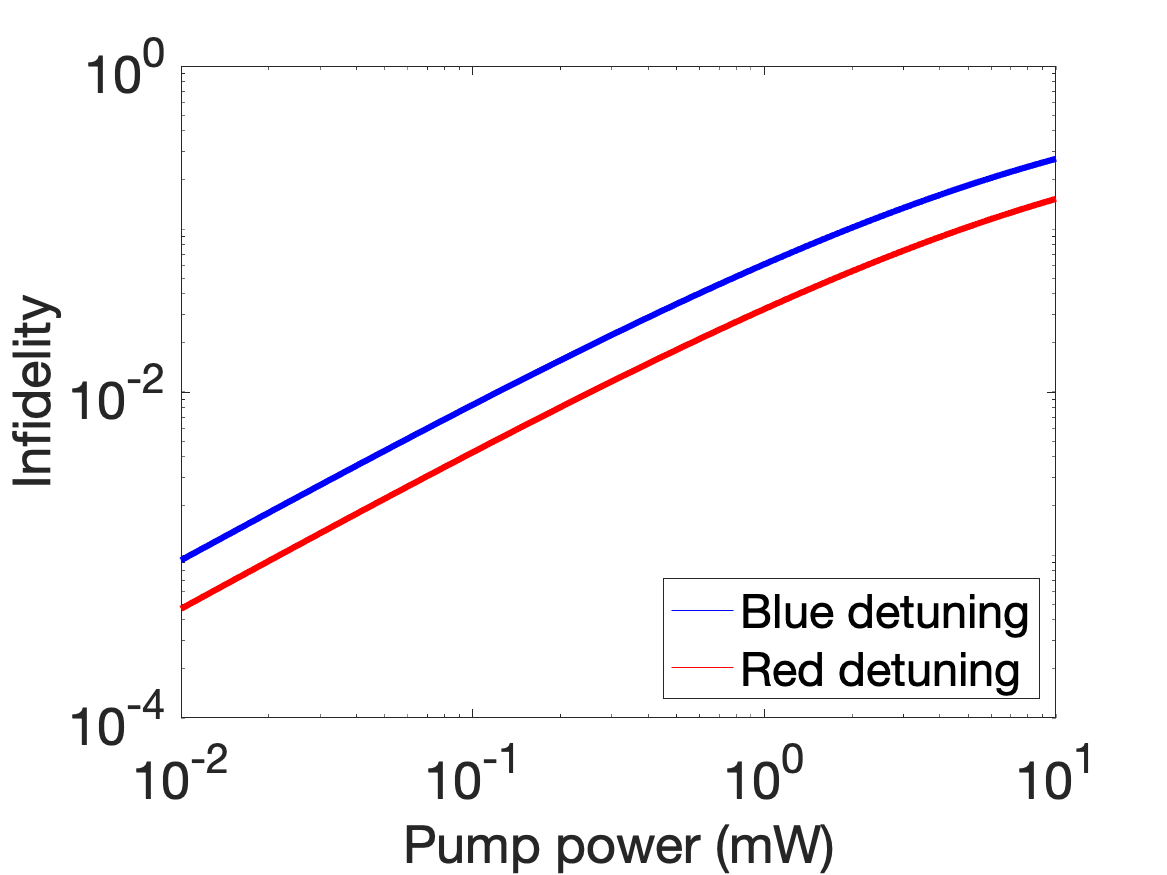}
  \caption{}
 \label{fig:infidelity}
 \vspace{-0.5cm}
\end{subfigure}
    \caption{Quantum transduction parameters vs. pump power, evaluated with respect to the RF quality factor (Q). \\(a) Efficiency; (b) Cooperativity; (c) Infidelity of heralded entanglement between two remote quantum devices.}
\end{figure}

We have characterized efficiency, cooperativity, and infidelity of quantum transduction against the laser pump power and the RF quality factor (Q). In \rfig{fig:transduction_efficiency} we observe that hybrid devices with higher Q achieve maximum conversion efficiency at lower pump powers. Also, the cooperativity (\rfig{fig:cooperativity}) scales linearly with respect to the quality factor and the pump power. We have finally simulated the infidelity of generating entangled states in two remote quantum devices in the blue- and red- detuned regimes (\rfig{fig:infidelity}), referring to the scheme in \cite{Loncar}. A larger pump power increases the probability of generating more microwave photons in the resonators, reducing therefore the fidelity of entanglement between two remote quantum devices. Increasing the Q by one order of magnitude in the microwave cavity improves the cooperativity, and reduces the infidelity by one order of magnitude as well. This result is important if we consider that 3D devices have demonstrated very slow heating rates under pump excitation at milli-Kelvin temperatures \cite{Rueda1}.
 
%($\SI{1.1}{} \text N_{\text{out}}$ per second, where $ \text N_{\text{out}}$ is the number of noises added output photons) \cite{Rueda1}.

\vspace{-0.7cm}

\section*{Acknowledgements}
This manuscript has been authored by Fermi Research Alliance, LLC under Contract No. DE-AC02-07CH11359 with the U.S. Department of Energy, Office of Science, Office of High Energy Physics. This work is funded by the Fermilab’s Laboratory Directed Research and Development (LDRD) program.

This research used resources of the U.S. Department of Energy, Office of Science, National Quantum Information Science Research Centers, Superconducting Quantum Materials and Systems Center (SQMS) under contract number DE-AC02-07CH11359. The NQI Research Center SQMS contributed by supporting the design of SRF cavities and access to facilities.

\/*
\section{Equation}
In entanglement generation, the target state is $|\psi>=c_1|01>+c_2|10>$.

In the blue detuning case, the photon generation follows a Poissonian process with photon generation rate $r_0$. Therefore, the probability that a photon is generated in one microwave cavity is
\begin{equation}
    P_1= r_0 \Delta t \exp(-r_0 \Delta t)
\end{equation}
The probability that no photon is generated is 
\begin{equation}
    P_0= \exp(-r_0 \Delta t)
\end{equation}
The probability that one photon is generated in each microwave cavity is
\begin{equation}
    P_{11}= P_1^2
\end{equation}
The probability that more than one photon is generated in each microwave cavity is
\begin{equation}
    P_{mn}= 2(1-P_0-P_1)
\end{equation}
The final state is $|\psi_f>=P_{00}|00>+P_{10}|10>+P_{01}|01>+P_{11}|11>+P_mn|mn>$. The infidelity is $P_{mn}+P_{11}$.

In the red detuning case. The probability that no photon is generated is 
\begin{equation}
    P_0= 1-\exp(-r_0 \Delta t)
\end{equation}
The final state is $|\psi_f>=P_{00}|00>+P_{10}|10>+P_{01}|01>+P_{11}|11>$. Since there cannot be more than one photons generated in each microwave cavity, the infidelity corresponds to the probability that the generated state is $|11>$: $P_{11}=(1-P_0)^2$.

*/

\end{document}